# Tunable Schottky barrier and high responsivity in graphene/Si-nanotip optoelectronic device

*Antonio Di Bartolomeo[1*], Filippo Giubileo[2], Giuseppe Luongo[1], Laura Iemmo[1], Nadia Martucciello[2], Gang Niu[3], Mirko Fraschke[4], Oliver Skibitzki[4], Thomas Schroeder[4,5], and Grzegorz Lupina[4]*

[1] Physics Department "E. R. Caianiello", University of Salerno, via Giovanni Paolo II, 84084, Fisciano, Italy

[2] CNR-SPIN Salerno, via Giovanni Paolo II, 84084, Fisciano, Italy

[3] Electronic Materials Research Laboratory, Key Laboratory of the Ministry of Education & International Center for Dielectric Research, Xi'an Jiaotong University, Xi'an 710049, China.

[4] IHP Microelectronics, Im Technologiepark 25, 15236 Frankfurt (Oder), Germany

[5] Institute of Physics and Chemistry, BTU Cottbus-Senftenberg, Konrad Zuse Str. 1, 03046 Cottbus, Germany

**ABSTRACT**

We demonstrate tunable Schottky barrier height and record photo-responsivity in a new-concept device made of a single-layer CVD graphene transferred onto a matrix of nanotips patterned on n-type Si wafer. The original layout, where nano-sized graphene/Si heterojunctions alternate to graphene areas exposed to the electric field of the Si substrate, which acts both as diode cathode and transistor gate, results in a two-terminal barristor with single-bias control of the Schottky barrier. The nanotip patterning favors light absorption, and the enhancement of the electric field at the tip apex improves photo-charge separation and enables internal gain by impact ionization.

These features render the device a photodetector with responsivity (3 $A/W$ for white LED light at 3 $mW/cm^2$ intensity) almost an order of magnitude higher than commercial photodiodes. We extensively characterize the voltage and the temperature dependence of the device parameters and prove that the multi-junction approach does not add extra-inhomogeneity to the Schottky barrier height distribution.

This work represents a significant advance in the realization of graphene/Si Schottky devices for optoelectronic applications.

## 1. INTRODUCTION

Graphene/silicon (Gr/Si) heterojunctions are key elements of many graphene-based devices such as photodetectors[1-3], solar cells[4-6], chemical-biological sensors[7-9], and high frequency transistors[10-14]. Such heterostructures are gaining interest from the semiconductor industry also for the potentiality to replace ultra-shallow doped junctions in modern complementary-metal-oxide-semiconductor (CMOS) technologies.

Despite recent progress in the deposition of Si layers on graphene[15], highest quality Gr/Si junctions are still formed by transferring large area graphene onto clean high-quality surfaces of Si single crystals[16]. Direct formation of graphene onto Si would be of higher technological relevance in the view of future applications, but this remains very challenging due to the formation of Si carbides. Here, we apply the graphene transfer technique to implement a new concept of Gr/Si photodiode with graphene on nano-patterned Si surfaces. We demonstrate that devices with graphene on Si nano-tip arrays are more performant than their large area, planar counterparts.

The zero-bandgap and linear energy-momentum relationship of graphene, which result in finite density of states, have been shown to enable energy Fermi level tuning and hence Schottky barrier height control by a single anode-cathode bias[17]. Adding an electrostatic gate can further improve the barrier control in a three-terminal barristor (variable barrier device)[11]. In our approach, the coexistence on the same graphene layer of junction areas with much bigger graphene regions exposed to the field of the substrate, which acts as well-coupled back-gate especially near the tips, enables improved control of the Schottky barrier height by a single applied bias. This peculiarity makes our device an effective two-terminal barristor with linear control of the barrier height. More importantly, while preserving the barrier uniformity, the nano-textured surface enhances light collection due to multiple reflections and the tip-enhanced field favors photo-charge separation with internal gain due to impact ionization. These features result in record responsivity, which is one to two orders of magnitude higher than in planar Gr/Si junctions[1] and about one order of magnitude better than in commercial semiconductor photodetectors.

## 2. EXPERIMENTAL DETAILS

Si-tip arrays (Fig. 1a) were prepared on degenerately doped ($\sim 10^{18}\ cm^{-3}$) n-type Si wafers. Fabrication of the Si-tip array includes a $SiO_2/Si_3N_4$ hardmask with photo-resist patterned by lithography, reactive ion etching (RIE) of Si, plasma-enhanced chemical vapor deposition (PECVD) of a thick $SiO_2$ layer completely covering the formed Si-tips, and a chemical-mechanical planarization step to reduce the $SiO_2$ thickness till revealing circular Si-tips surface of given diameter. Further details on the fabrication process can be found elsewhere[18]. Just before the graphene transfer, the Si-tip substrates were dipped in a 0.5% hydrofluoric acid solution for 10 seconds to remove only the native $SiO_2$ on Si-tips[19-20] and enable formation of clean Gr/Si junctions. Graphene was transferred from commercially available Cu foils using a wet transfer process[21]. Immersion in deionized water and subsequent dry process by nitrogen-gas blowing helped in H-passivating the surface dangling bonds. Fig. 1b) shows a scanning electron microscopy (SEM) image taken after graphene transfer: Five Si-tips with a diameter of about 50 nm are seen underneath the graphene layer and wrinkles, characteristic of CVD graphene grown on Cu, can be clearly identified. Fig. 1c) shows a SEM cross-section of one of the Si-tips with graphene.

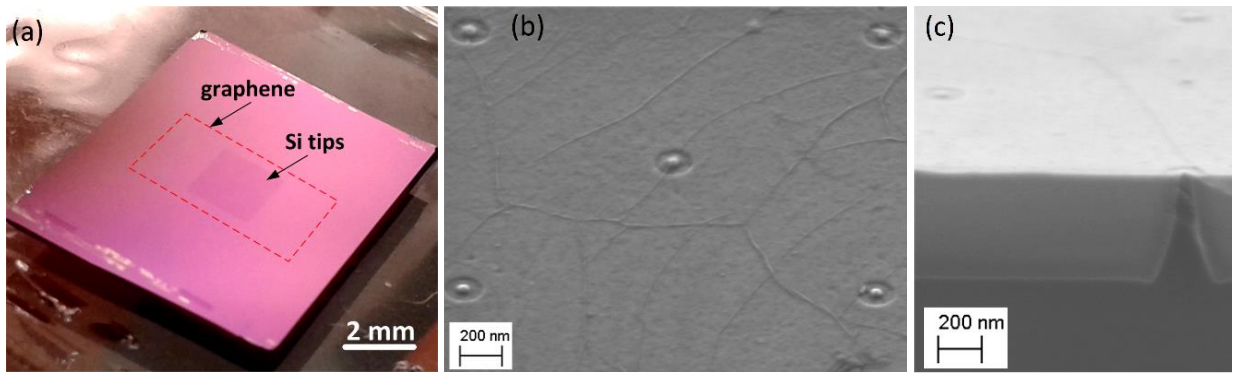

**Figure 1.** a) Photograph of a single chip with a $2.5 \times 2.5\ mm^2$ Si-tip array covered by a monolayer graphene transferred from Cu. b) SEM top view of the tips emerging from the SiO$_2$ insulating layer and covered by graphene. The diameter of the emerging tips is $\sim 50\ nm$. Graphene wrinkles characteristic of graphene transferred from Cu are visible. c) SEM cross-section image showing one of the Si-tips embedded in SiO$_2$ and covered by graphene. The tip pich size is $1.41\ \mu m$ and the tip height is $\sim 0.5\ \mu m$

To evaluate the quality of graphene, Raman spectroscopy measurements with a $514\ nm$ laser source (spot size $\sim 600\ nm$) were performed. Fig. 2a) shows a representative spectrum taken from the area between the Si-tips. Beside the characteristic 2D and G bands, a typical feature of monolayer graphene, a very small D peak related to defects is seen at about 1350 cm$^{-1}$. As shown by peak intensity mapping measurements presented in Fig. 2b), the intensity of the D peak does not correlate with the positions of the Si-tips. The observed local increase in the D band intensity is most probably related to the presence of multilayer graphene islands[22]. Similarly, no particular correlation between the 2D/G peak intensity ratio and the positions of the Si-tips was revealed (Fig. 2(c)). Another small peak appears at 2450 cm$^{-1}$, generally indicated as D+D", and interpreted as a combination of D and D" phonons, the latter belonging to the in-plane longitudinal acoustic branch[23].

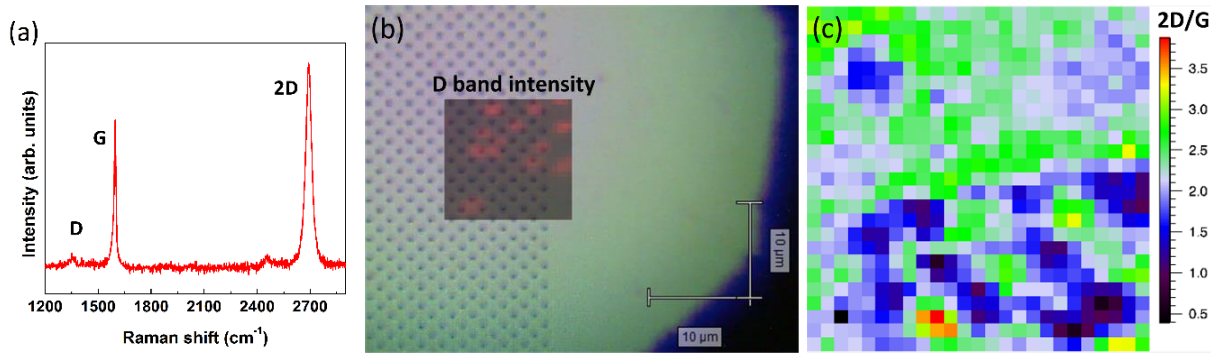

**Figure 2.** a) Representative µRaman spectrum taken after graphene transfer. b) D band intensity distribution extracted from Raman mapping measurement on an area of $12 \times 12\ \mu m^2$ overlaid on optical image. c) 2D/G intensity map taken from the same $12 \times 12\ \mu m^2$ area indicated in b).

Sheet resistance of the graphene layer measured using 4-point technique beyond the Si-tip array was $\sim 0.9\ k\Omega/\square$, a value in the range typically reported for CVD graphene on Cu[24].

The setup used for electrical measurements of the Gr/Si heterojunction is illustrated in Fig. 3a). The top graphene layer was contacted with evaporated Au, while ohmic contact with the scratched bottom Si substrate was made with Ag paste. Electrical measurements were performed in a Janis probe station with pressure and temperature control. The top-injection configuration was adopted, with the biasing lead on graphene and the Ag electrode grounded. The measurements were carried out at atmospheric pressure.

## 3. RESULTS AND DISCUSSION

The dark I-V characteristics of the Gr/Si-tip heterojunction in the temperature range 120-390 K are shown in Fig. 3b). The device exhibits a rectifying behavior with the forward current at positive bias, as expected for p-type graphene on n-Si. The p-type doping is usually observed in air-exposed graphene[10,25]. The current for a given voltage increases with rising temperature, which is typical of thermionic emission in this kind of devices. At modest positive bias, the room and higher temperature forward I-V curves show an almost ideal diode behavior. At lower temperatures, namely $T < 250\ K$, a new feature appears in the lower bias part of the forward I-V curves, where the current is dominated by a leakage component that adds non-linearity to the semi-log I-V plot. This leakage component is usually attributed to generation and recombination of carriers in the charge space region, field emission and thermionic field emission or surface/edge effects that may lead to local barrier lowering [26-27]. Such component becomes relevant when the low-temperature suppresses the thermionic emission, e.g. at $T = 121\ K$, where it manifests on the interval $0 < V < 0.25\ V$.

To gain insight into carrier transport across the Gr/Si-tip device, we focus on the I-V curve at $T = 300\ K$. Fig. 3c) shows the measured data together with the best fitting curve as predicted by the ideal Schottky model[16,28-29], which is described by the following I-V relation:

$$I = I_0 \left[ e^{q(V-R_sI)/nkT} - 1 \right] \quad (1)$$

with

$$I_0 = AA^*T^2 e^{-\Phi_B/kT} \quad (2)$$

where $I_0$ is the reverse saturation current, $A$ the contact area, $A^* = 4\pi q m^* k^2/h^3$ the effective Richardson constant with $m^*$ the electron effective mass, $T$ the absolute temperature, $\Phi_B$ the Schottky barrier height, $k$ the Boltzmann constant, $q$ the electronic charge, $n$ the ideality factor that takes into accounts possible deviations from the thermionic regime, and $R_s$ the series resistance. More specifically, $R_s$ is the lump sum of bulk Si, graphene, metal leads and contact resistances, and is dominated by graphene.

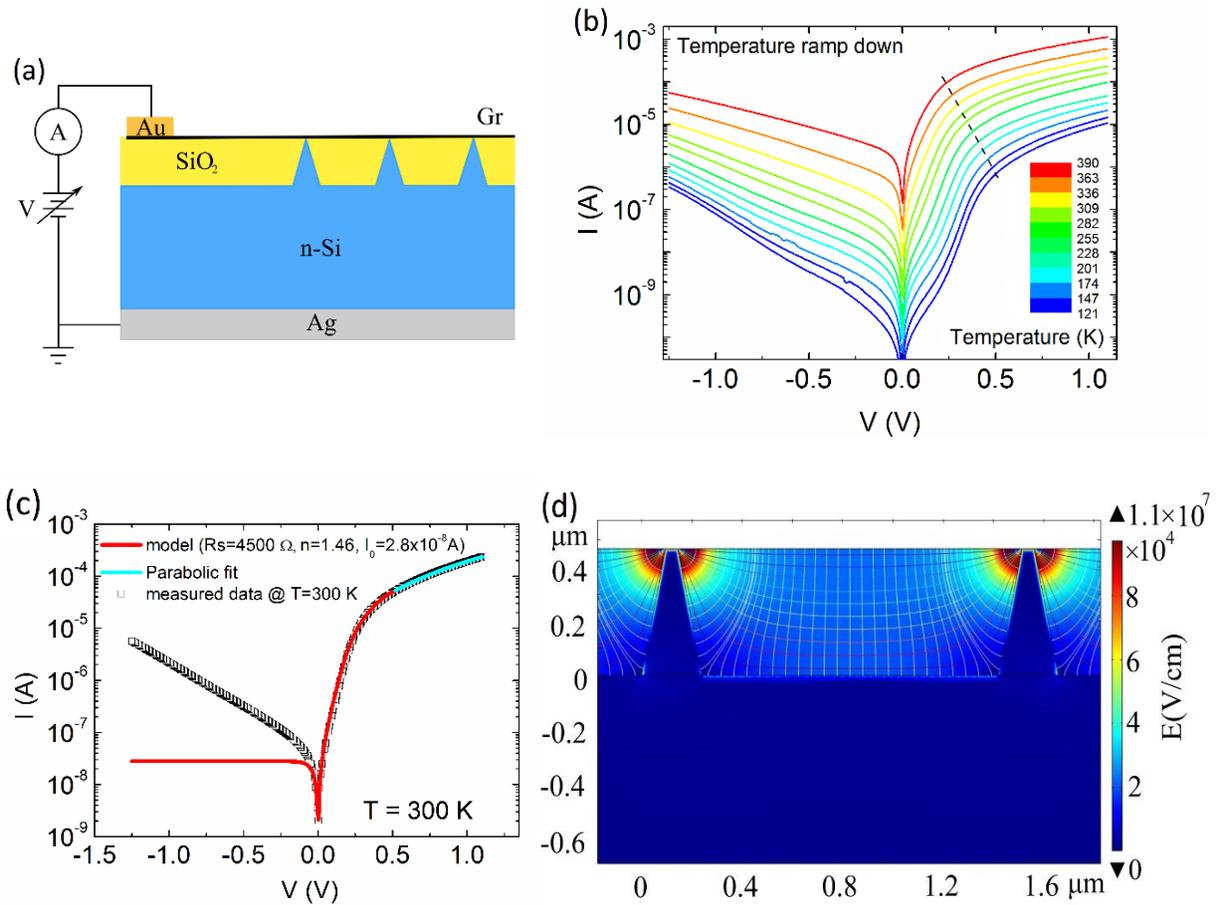

**Figure 3.** a) Layout and measurement setup of the Gr/nSi-tip device. b) I-V characteristics in the temperature range 120 to 390 K and in steps of $\sim 30\,K$ measured in dark and at atmospheric pressure. c) I-V curve measured at $T = 300\,K$ (black squares) and fit of the ideal Schottky diode model of eq. (1) (red line). In flat-band condition, that is for $V > 0.5\,V$, the I-V curve is better

described by a SCLC model, $I \sim V^2$ (cyan line). d) Numerical simulation (by COMSOL software) of the electric field between two Si-tips under reverse bias (-1V), showing field amplification (cyan to reddish color) near the tip edge, where the gating effect of the substrate is more effective. Field (white) and equipotential (multi-color) lines are shown.

As seen in Fig. 3c), eq. (1) provides a perfect fit in the range $0 < V < 0.5\ V$; at higher bias, the Gr/Si-tip diode enters a high injection regime, where the voltage drop across the series resistance strongly limits the exponential increase of the current, until the barrier reaches the flat-band condition and the I-V characteristic is dominated by the series resistance. In this region, henceforth referred as flat-band, the current is better described by a quadratic relation, $I \sim V^2$, typical of space charge limited conduction (SCLC). The gating effect of the substrate increases the p-doping of graphene when $V > 0\ V$ and this is the origin of the quadratic dependence. Indeed, in flat-band condition, $V \approx R_s I$, and the graphene-dominated $R_s$ is proportional to the inverse bias ($R_s \sim (q n_q)^{-1} \sim (CV)^{-1}$ where $n_q$ is the graphene carrier density and $C$ the gate capacitance per unit area), which makes the current to scale as $V^2$. The gating effect of the substrate is particularly effective around the tips, where the electric field is stronger, as shown in Fig. 3d).

Fig. 3c) also evidences that, in reverse bias ($V < 0\ V$), the current dramatically deviates from the constant behavior predicted by eq. (1), implying that it is not the usual saturation current of the diode. Since $\ln I$ increases linearly with $|V|$, the common modeling by a simple shunt resistance seems not appropriate in this case. A better explanation is provided by a bias dependence of the Schottky barrier, due to a combination of image force lowering and low density of states of graphene in the absence of Fermi level pinning, as we will discuss in the following.

Fig. 4 shows the rectification factor $r$, the series resistance $R_s$ and the ideality factor $n$ over the explored temperature range. The rectification factor $r$ is here defined as the ratio of the on/off current at $V = \pm V_{FB}$, $I(V_{FB})/I(-V_{FB})$, where $V_{FB}$ is the voltage corresponding to the onset of the flat-band region (that is the region to the right of the dashed-black line in Fig. 3b). Both r and $V_{FB}$ decrease with increasing temperature. In particular, $r$, which is ~120 at room temperature, has a monotonic behavior with two possible regimes crossing at $T \sim 260\ K$. Below this temperature, there is a reduced rate $dr/dT$, likely correlated with the deviation from the pure thermionic emission.

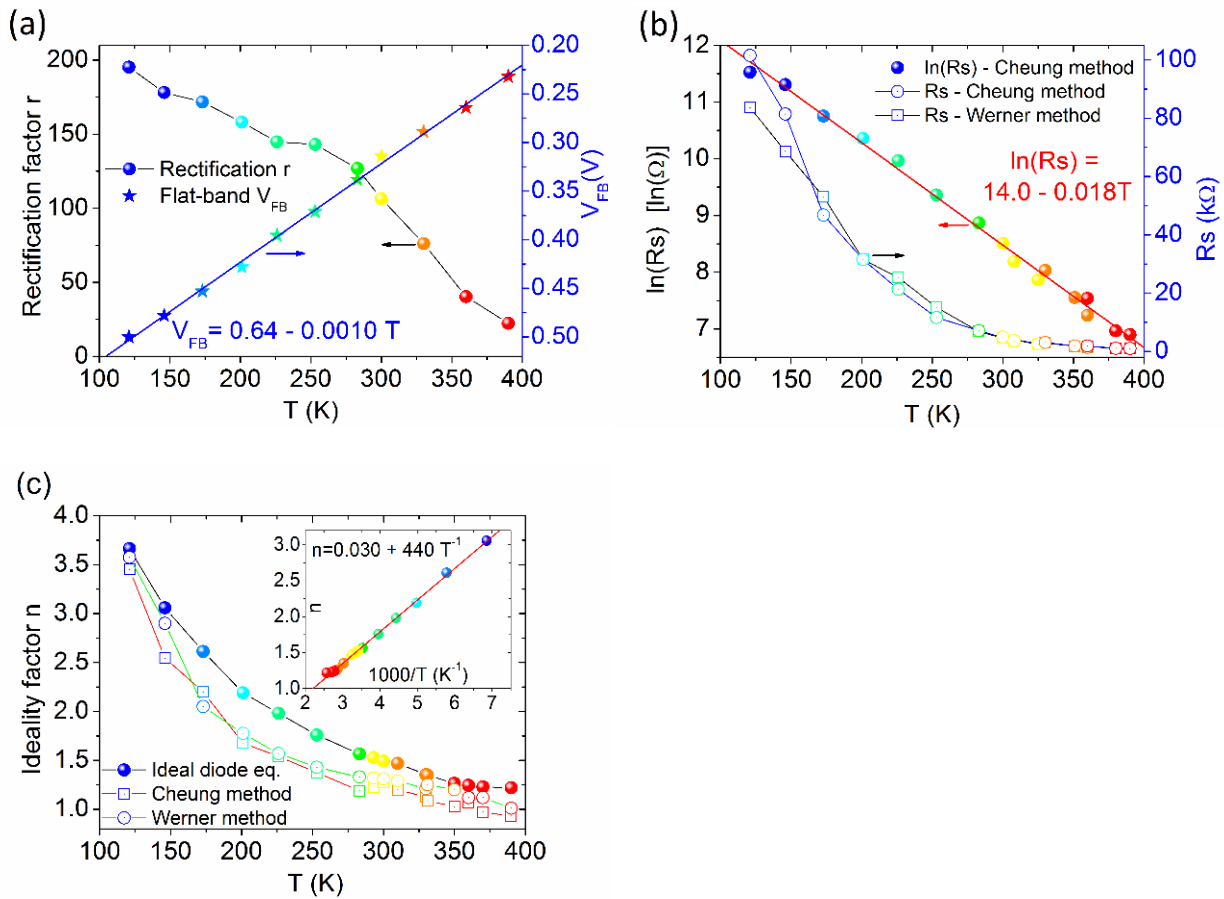

**Figure 4.** a) Rectification factor $r$ and flat-band voltage $V_{FB}$ (in reverse scale) as a function of temperature $T$. The ideality factor is defined as $r = I(V_{FB})/I(-V_{FB})$, where $V_{FB}$ is the voltage

corresponding to the flat-band condition. b) Series resistance $R_s$ (right scale) and $\ln R_s$ (left scale) as a function of the temperature $T$, obtained from Cheung and Werner methods. c) Comparison of the temperature dependence of the diode ideality factor $n$ extracted from ideal diode (full circles), Cheung (empty squares) and Werner (empty circles) methods. The inset shows the linear dependence of $n$ on $T^{-1}$.

$V_{FB}$ linearly decreases with temperature with a slope $|dV_{FB}/dT| \approx 1\ mV/K$, a behavior similar to the forward voltage drop of a typical *pn*-diode. As the temperature increases, the flat-band condition, $V \approx R_s I(V)$, is reached at lower bias due to the strong $I(T)$ dependence, as described by eq. (2). Indeed, the linear behavior of $V_{FB}$ implies a strongly decreasing $R_s(T)$ to counterbalance the exponential growth of $I(T)$ with temperature.

We used I-V data below and around $V_{FB}$, and followed the Cheung method[30-31] to extract $R_s$ and $n$ at a given temperature from the slope and the intercept of the corresponding $dV/d(\ln I)$ vs. $I$ plot, since

$$\frac{dV}{d(\ln I)} = \frac{nkT}{q} + R_s I. \quad (3)$$

$R_s$ and $n$ were also obtained from $(dI/dV)/I$ vs. $dI/dV$ plots (Werner method[32]), considering that

$$\frac{dI/dV}{I} = \frac{q}{nkT}\left[1 - R_s\left(\frac{dI}{dV}\right)\right] \quad (4)$$

Both eq. (3) and (4) are valid in the limit of $V \geq 3nkT/q$. The consistent results are shown in Fig. 4b) and 4c). Remarkably, $R_s$ exhibits the expected exponential decrease with $T$ $(R_s \sim \exp(-\alpha T))$, that is a semiconductor behavior with negative temperature coefficient of resistance, $dR_s/dT$. Neither Si[33,34] nor metals or ohmic contacts[35] can account for the negative $dR_s/dT$ in the temperature range under investigation. The semiconductor behavior can only originate from graphene. Indeed, the resistivity of graphene, especially at the lower carrier density close to the Dirac point and in the presence of defects or impurities, has been reported to decrease with rising temperature on exfoliated or CVD monolayer graphene, both suspended[36] or deposited on substrate[37-39]. A semiconductor behavior has been reported also for bilayer and few layer graphene on substrate over a wide temperature range[40-42]. The negative temperature coefficient in graphene is the result of competing mechanisms, with the thermally activated transport through inhomogeneous electron-hole puddles as the main recognized cause[39]. CVD graphene is more vulnerable to impurities or charged defects during the transfer process, and is particularly prone to develop electron-holes puddles that tend to produce a negative $dR_s/dT$.

The ideality factor was further estimated from the slope of straight-line fitting the thermionic part of the $\ln(I)$ vs. $V$ plot (i.e. the part between 0 and $\sim V_{FB}$, after excluding the leaky portion at lower biases), which according to eq. (1) and for $V \gg R_s I$ is described by

$\ln I = \ln I_0 + \frac{q}{nkT} V.$ (5)

While the temperature dependence of $n$ is the same (Fig. 4c), eq. (5) provides values which are 10 to 20 % higher than those obtained with Cheung and Werner methods. The slope of a fitting straight line is usually more accurate than the intercept for the estimation of diode parameters, so the method based on eq. (5) is considered more reliable. $n$ decreases with increasing temperature ($n = c + T_0/T$ with $c$ and $T_0$ constants, as shown in the inset of Fig. 4c) and approaches the ideal value of 1 at the highest temperatures. This behavior confirms that the thermionic emission is the dominant carrier conduction process at high temperatures. On the contrary, at low temperatures, the growing $n$ indicates that other transport phenomena, as generation-recombination in the space charge region or thermionic field emission, add to thermionic emission. Furthermore, the observed temperature dependence of the ideality factor is a signature of Schottky barrier spatial inhomogeneity and of deformation of the barrier when a bias voltage is applied[43].

According to eq. (2), a plot of $ln(I_0/T^2)$ vs. $1/T$ (Richardson plot) is a straight line, whose slope and intercept are used to evaluate the Schottky barrier height $\Phi_B$ and $ln(AA^*)$ at a given bias:

$$ln\left(\frac{I_0}{T^2}\right) = ln(AA^*) - \frac{\phi_B}{k}\frac{1}{T}. \qquad (6)$$

In reverse bias, when $e^{q(V-R_sI)/nkT} \ll 1$, $I_0(T)$ is directly measured on the curves of Fig. 3b). At zero bias, $I_0$ is extrapolated to $V = 0\ V$ as the intercept of the straight line fitting the thermionic part of the forward I-V characteristics. In forward bias, when $e^{qV/nkT} \gg 1$ and $V \gg R_sI$, eq. (1) and (2) combine to yield a slightly modified Richardson equation, that requires $n$ to estimate the Schottky barrier height:

$$ln\left(\frac{I_0}{T^2}\right) = ln(AA^*) - \frac{\phi_B - V/n}{k}\frac{1}{T}. \qquad (7)$$

Examples of Richardson plots for a subset of applied biases are shown in Fig. 5a), while a more complete set of measured $\Phi_B(V)$ is summarized in Fig. 5b). Remarkably, $\Phi_B(V)$ exhibits a linear increase with $V$, and has the value $\Phi_B \approx 0.36\ eV$ at zero bias. Fig. 5b) depicts $ln(AA^*)$ with a very weak dependence on $V$. Considering the average value $ln(AA^*) \approx -16$ and an effective junction area of $6.079 \times 10^{-5} cm^2$ (estimated from the number of tips and conservatively assuming a circular junction area with mean radius of 25 nm), the Richardson constant results $A^* = 0.002\ A/(K^2 cm^2)$, a value significantly lower than the theoretical $112\ A/(K^2 cm^2)$ for electrons in Si. Similar low values, ranging from $10^{-3}$ to $10^{-1} A/(K^2 cm^2)$ have been reported for Gr/Si planar heterojunction[44-46] and revised Schottky diode equations, based on Landauer formalism[45] or including the massless Dirac fermion nature of carriers in graphene[47-48], have been proposed to explain this anomaly.

The low value of $A^*$, the temperature dependence of $n$, the linear bias dependence of $\Phi_B$ as well as the deviation from linearity of the Richardson plot at low temperatures, can be ascribed to spatial inhomogeneities in the Schottky barrier height[43], as already mentioned. Since the Gr/Si-tip device under study is made of more than $3 \times 10^6$ nanojunctions, minimal tip-to-tip variation could result in significant barrier height fluctuations. Hills and valleys in the barrier-height distribution can be caused e.g. by local effective barrier lowering due to field emission from tips with narrower radius of curvature or by surface defects or contaminates, possibly introduced during the fabrication process.

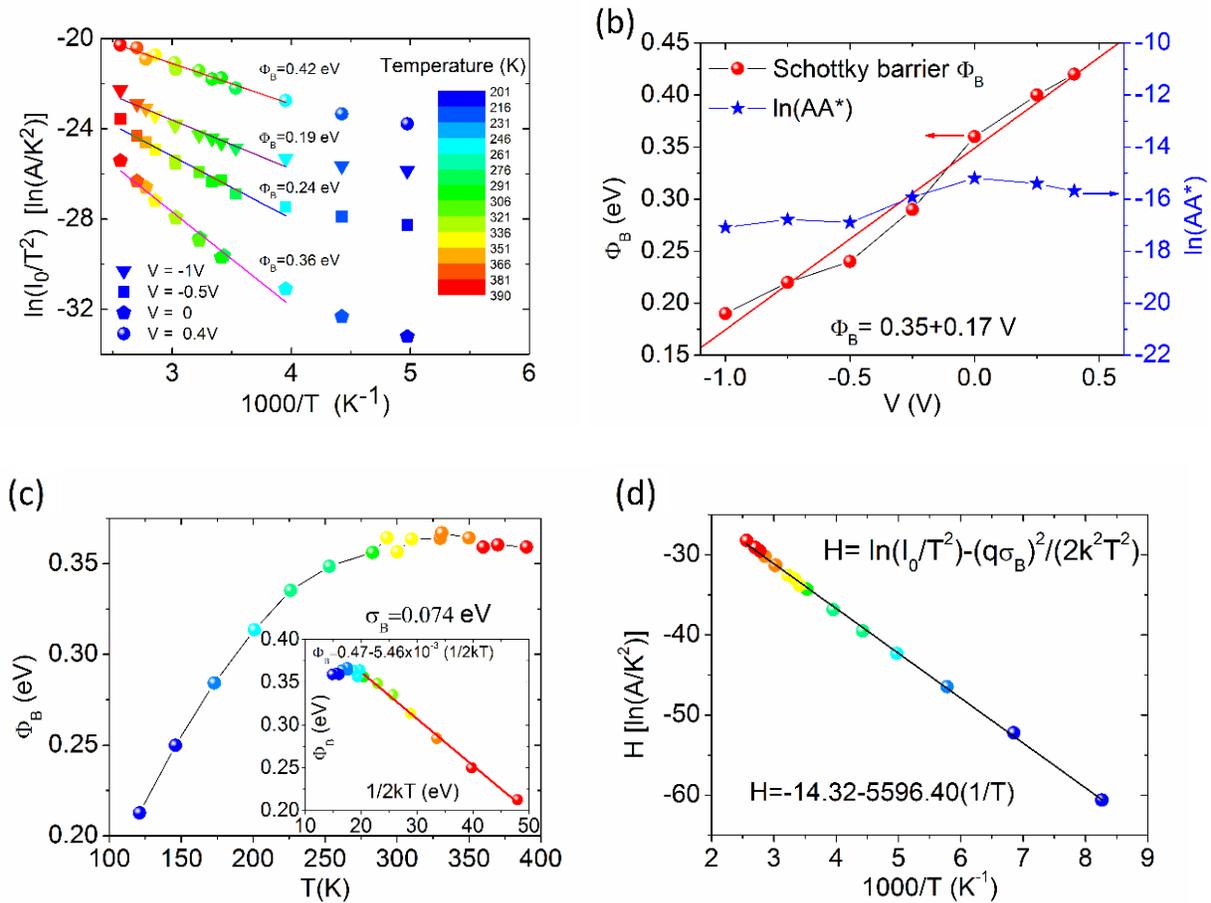

**Figure 5.** a) Plot of $ln(I_0/T^2)$ vs. $1000/T$ (Richardson plot) with linear fit to extract the Schottky barrier height $\Phi_B$ at different bias V. Richardson plots show non-linearity at low temperature due

to Schottky barrier inhomogeneities and deviation from the pure thermionic emission theory; accordingly, to estimate $\Phi_B$, the two lowest temperature points were excluded from the fit. b) Dependence of Schottky barrier height $\Phi_B$ and $\ln(AA^*)$ on $V$. c) $\Phi_B$ at zero bias as a function of temperature; the inset shows the fit of eq. (8) from which a value of the Schottky barrier inhomogeneity $\sigma_B \approx 74$ meV is obtained. d) Modified Richardson plot according to the Gaussian distribution of the Schottky barrier height (eq. (9)).

An applied bias deforms the lower and higher barrier patches, causing $\Phi_B(V)$ dependence[43]. Indeed, a major disadvantage of the tip approach could be the limited shape control. Rather than a simple Si(001) plateau facet region, a tip could be a multi-facetted round shape Si structure. Different Si facets might have different work function values and further contribute to Schottky barrier inhomogeneities. Hence, to check the Schottky barrier spatial distribution, we calculated $\Phi_B(T)$ using eq. (2) with the extrapolated values of $I_0(T)$ at zero bias, and studied its temperature dependence (Fig. 5c).

The decreasing barrier height with lowering temperature is easily understood considering that the current becomes gradually dominated by electrons able to surmount the lower barrier patches, and this gives a reduced apparent barrier height. Assuming a Gaussian spatial distribution for $\Phi_B$, with mean $\Phi_{Bm}$ and the standard deviation $\sigma_B$, the temperature dependence of the measured (apparent) barrier height $\Phi_B$ at zero applied bias is expected to follow the relation[43]:

$$\Phi_B = \Phi_{Bm} - \frac{q\sigma_B^2}{2kT}. \qquad (8)$$

The standard deviation $\sigma_B$ is a measure of the inhomogeneity of the Schottky barrier height (the lower $\sigma_B$ the more uniform is $\Phi_B$), and, according to eq. (8), can be extracted from the plot of $\Phi_B$ vs. $\frac{1}{2kT}$, as shown in the inset of Fig. 5c). As already mentioned, the Schottky barrier distribution can be deformed by applied bias. In particular, Werner et al.[43] have demonstrated that a linear $\Phi_B(V)$ implies a voltage independent ideality factor and that a linear $n$ vs. $1/T$ results from the narrowing of the Gaussian barrier distribution (i.e. of $\sigma_B$) with forward bias, meaning that the application of a forward bias homogenize the barrier fluctuations. The value of $\sigma_B \sim 74\ meV$ at zero bias is in agreement or below what has been reported for planar Gr/semiconductor heterojunctions[44,46,49]. This lead to the remarkable result that the tip-geometry and the transfer process do not introduce extra-inhomogeneity. However, we point out that the measured $\sigma_B$, despite its low value, is still enough to affect the low temperature part of the I-V characteristics and the Richardson plots.

As consistency check, we notice that $\Phi_{Bm} = 0.47\ eV$ extracted from eq. (8) is within 1.5 $\sigma_B$ from the apparent $\Phi_B$ at high temperature estimated using eq. (2). Together, eq. (2) and (8), give

$$H \equiv \ln\left(\frac{I_0}{T^2}\right) - \frac{q^2 \sigma_B^2}{2k^2 T^2} = \ln(AA^*) - \frac{q\Phi_{Bm}}{kT}, \quad (9)$$

which suggests a modified Richardson plot of $H$ vs $1000/T$ from which a more accurate and higher value of $A^* = 0.015\ A/(K^2 cm^2)$ can be estimated, while $\Phi_{Bm} = 0.48\ eV$ remains practically unaffected. We remark here that $A^*$ is possibly underestimated, given our conservative assumption of constant full-contact area between Gr and Si-tips; we also underline that $\Phi_{Bm}$ is consistent with other reported evaluations[11,50-51] and, as we discuss in the following, matches the prediction of Schottky-Mott theory.

The Schottky barrier height depends on the graphene work function, the Si electron affinity as well as on the interface states density and on the thickness of a possible interfacial layer of atomic dimension, often due to native oxide, that is transparent to electrons but able to withstand a potential drop[16,29,52]:

$$\Phi_B = \Phi_{Gr} - X_{Si} - q\Delta \quad (10)$$

where $\Phi_{Gr} = 4.5 \div 4.6\ eV$ is the work function of graphene[53-54], $X_{Si} = 4.05\ eV$ is the electron affinity of Si and $\Delta$ is the potential drop across the interfacial layer. High density of interface states at a given energy in the Si bandgap usually leads to pinning of the Fermi level and can result in a significant discrepancy from eq. (10). However, due to negligible interaction with chemically inert graphene, the formation of interface states is suppressed in graphene-semiconductor junctions if the Si-surface is defects-free and with saturated dangling bonds[55]. An ideal, trap free interface would result in unpinned Fermi level and yield a Schottky barrier height

$$\Phi_B = \Phi_{Gr} - X_{Si}, \quad (11)$$

(Schottky-Mott relation) in the range $0.45 \div 0.55\ eV$ for undoped graphene. The Fermi level unpinning enables easy modulation of the Schottky barrier, a feature that can be exploited to tune Gr/Si devices to match specific performance requests[11,56]. Deviations from the Schottky-Mott prediction are mainly due to image force lowering[57] or hot electrons barrier lowering[58-59]. Neglecting the field enhancement by the tip (values of the field up to $10^7$ V/cm can be achieved, as shown in Fig. 3d), the maximum built-in electric field at the Gr/Si junction[29], is

$$E_m = \sqrt{(2qN|\phi_i|)/\varepsilon_{Si}} \approx 3 \times 10^5\ V/cm \quad (12)$$

($N = 10^{18}\ cm^{-3}$ is the Si doping density, $\phi_i \approx 0.3\ V$ is the built-in potential and $\varepsilon_{Si} = 12\varepsilon_0$ is the dielectric constant of Si) which corresponds to a Schottky barrier lowering by image force:

$$\Delta\Phi_B^I = q\sqrt{qE_m/(4\pi\varepsilon_{Si})} = q\sqrt[4]{q^3 N|\phi_i|/(8\pi^2\varepsilon_{Si}^3)} \approx 0.06\ eV \quad (13)$$

When a bias is applied to the junction, $|\phi_i|$ is replaced by $|\phi_i - V|$ in eq. (12) and (13), and $\Delta\Phi_B^I$ is increased (decreased) by a reverse (forward) bias, as depicted in Fig. 6a. At zero bias, adding $\Delta\Phi_B^I$ from eq. (13) to the measured $\Phi_B \approx 0.36\ eV$ brings the Schottky barrier height close to the ideal Schottky-Mott value. This result confirms the good quality of the Gr/Si-tip interfaces in the device under study.

A good quality interface also excludes the chance of Fermi level pinning. In graphene, the Fermi energy, $E_F$, at room temperature, is approximately related to the free carrier density $n_q$ by a quadratic relation

$$n_q \approx n_{q0} + E_F^2/(\pi \hbar^2 v_F^2) \qquad (14)$$

($n_{q0}$ is the intrinsic carrier density and $v_F \approx 10^6 \ m/s$ is the Fermi velocity of graphene). Eq. (14) enables fine tuning of $E_F$ via $n_q$, which can be easily controlled by an electric field (electrostatic doping). In heterojunctions with unpinned Fermi level, $E_F$ modifies the graphene work function and hence the Schottky barrier height. P-doping of graphene increases $\Phi_B$ and n-doping decreases it, as displayed in Fig. 6a) for graphene on n-type Si. The electrostatic Schottky barrier variation, $\Delta\Phi_B^E = -E_F$, adds to the image force barrier lowering $\Delta\Phi_B^I$, to set $\Phi_B$ (Fig. 6a). Noteworthy, in the device under study, $E_F$ modulation by electrostatic doping is achieved via the electric field (Fig. 3d) generated by the same voltage used to bias the junction.

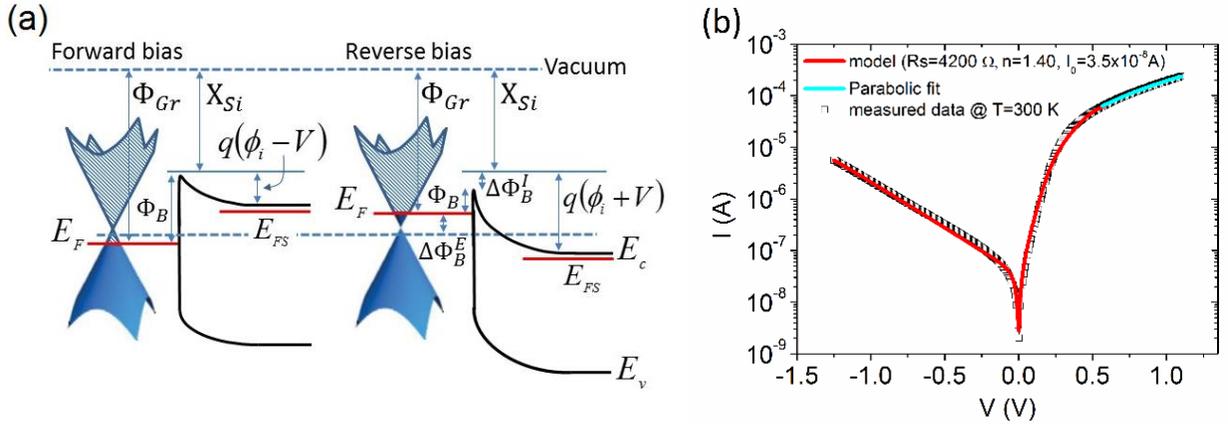

**Figure 6.** a) Band diagrams of the Gr/Si-tip junction in forward and reverse bias. $\mathrm{X}_{Si}$, $\Phi_{Gr}$ and $E_F$ designate the Si electron affinity, the graphene work function and Fermi level, respectively; $E_C$ / $E_V$ are the bottom/top of the conduction band and $E_{FS}$ the Fermi level of Si (the dashed blue line

represent the Fermi level at zero-bias, with an assumed low graphene p-doping). $\phi_i$ is the built-in potential and the $V$ applied bias. In forward (reverse) bias the change of the graphene Fermi level leads to an increase (decrease) of the Schottky barrier height ($\Delta\Phi_B^E = -E_F$). The graphene Fermi level shift, which is required to allocate more and more charge to mirror the bias-dependent immobile charge of the semiconductor depletion layer, is due to the low density of states of graphene around the Dirac point. Also shown is the variation of Schottky barrier height, $\Delta\Phi_B^I$, caused by image force. b) Fitting of eq. (17) to the I-V data measured at T=300 K.

Both $\Delta\Phi_B^I$ and $\Delta\Phi_B^E$ introduce a sublinear dependence on the applied bias V in the Schottky barrier height (roughly as the $\sqrt[4]{|V|}$ from eq. (13) and $\sqrt{|V|}$ from eq. (14), respectively, since $|\Delta\Phi_B^E| = |E_F| \sim \sqrt{qn} \sim \sqrt{V}$), which can be written as

$$\Phi_B(V) = \Phi_{B0} + \Delta\Phi_B(V). \quad (15)$$

$\Delta\Phi_B(V)$ is positive in forward bias and negative in the reverse bias.

The effect of electrostatic doping, $\Delta\Phi_B^E$, was first included in the diode eq. (1) and (2) by Tongay et al.[17] who obtained a reverse saturation current $I_0$ depending on the exponential of $\sqrt{|V|}$. A similar behavior was proposed with different approaches also by Sinha et al.[45] and Liang et al.[47-48]. However, for the Gr/Si-tip device under study, the measured barrier height of Fig. 5b) shows that $\Phi_B(V)$ is better described by a linear relation. Accordingly, we write $\Delta\Phi_B(V) = \gamma(V - R_s I)$, with $\gamma = 0.17\ eV/V$ taken from the straight-line fit of Fig. 5 (b), and

$$I_0 = AA^*T^2 e^{-[\Phi_{B0}+\gamma(V-R_s I)]/kT} \quad (16)$$

By defining $n = 1 - \gamma = 1 - \partial\Phi_B/\partial V$, eq. (1) and eq. (16) can be combined to obtain:

$$I = AA^*T^2 e^{-q\Phi_{B0}/kT} e^{-q(V-R_s I)/nkT}\left[1 - e^{-q(V-R_s I)/kT}\right] \quad (17)$$

which is a common way to rewrite the Schottky diode equation to phenomenologically include the ideality factor for both the forward and reverse current.

Fig. 6b) shows that eq. (17) provides an excellent fit to the measured data at 300 K over the whole bias range (the flat-band regime obeys a quadratic law, as explained before). The dependence of $I_0$ as the exponential of a linear rather than a sub-liner function of $|V|$ is likely due to the effect of Si substrate which, especially in the vicinity of the tips (Fig. 3d), acts as strongly-coupled gate, which would linearly shift $E_F$. Consequently, the Gr/Si-tip device behaves as a barristor, with linear control of the Schottky barrier height as for the device of Yang et al.[11], but without the need of a third gate electrode.

Another important effect which can lead to a stronger V-dependence of the reverse current is the Schottky barrier lowering caused by hot electrons[58-59] that might originate even a quadratic $\Delta\Phi_B(V)$. In this scenario, the gating effect induces abrupt band bending around the Schottky barrier that increases the lateral field, which in turn produces significant enhancement of hot carriers.

We also studied the photoresponse of the Gr/Si-tip device by shining light from the top, on the graphene layer. Fig. 7a) compares the I-V curves obtained in darkness and under $3 \ mW/cm^2$ white LED light; it shows clear photocurrent in reverse bias, and photovoltaic effect with 60 nA short circuit current (a factor ~50 higher than the dark current at zero bias) and 70 mV open circuit voltage. In reverse bias, the Gr/Si-tip device can be used as a photodetector: The electron-hole pairs generated by incident photons in the n-Si space-charge region and, in minimal part, in graphene are separated by the strong tip-enhanced electric field, leading to a photocurrent[20]. The inset of Fig. 7a) shows the stable photoresponse when the light is switched on and off, at a bias of -0.5 V, corresponding to a responsivity $\mathcal{R}_{ph} = I_{ph}/P_0 = (I_{ligth} - I_{dark})/P_0 \approx 3 \ A/W$ (here, $I_{ph}$ is the photocurrent and $P_0$ the incident optical power). Fig. 7b) shows the photoresponse, at the same bias of -0.5 V, to the near IR radiation produced by a 880 nm LED. The Gr/Si-tip device is expected to show high sensitivity around this wavelength, since Si has an absorption peak[3] at ~930 nm. The current steps of Fig. 7b) and of its top inset are the response of the Gr/Si-tip device to increased input electrical power to the irradiating diode: The photocurrent displays a monotonic growth with IR intensity (as seen also in the bottom insets of the figure). The radiation intensity in $W/cm^2$ reaching the Gr/Si-tip junction was measured to be ≤1% of the IR diode supply power. Hence, Fig. 7b) shows that the minimum detectable IR intensity by the Gr/Si-tip heterojunction is less than 100 $\mu W/cm^2$ and its responsivity is $\mathcal{R}_{ph} \geq 0.3\ A/W$ at intensity < 1 $mW/cm^2$. The measured responsivity is one or two orders of magnitude higher than the values reported to date for graphene/Si planar-junctions[3,60-63] (with maximum of 225 $mA/W$ at 488 $nm$)[1]. The graphene/Si-tip device appears also competitive when compared to semiconductor photodiodes on the market, whose typical responsivity is around 0.5 $A/W$, both for visible and near IR light.

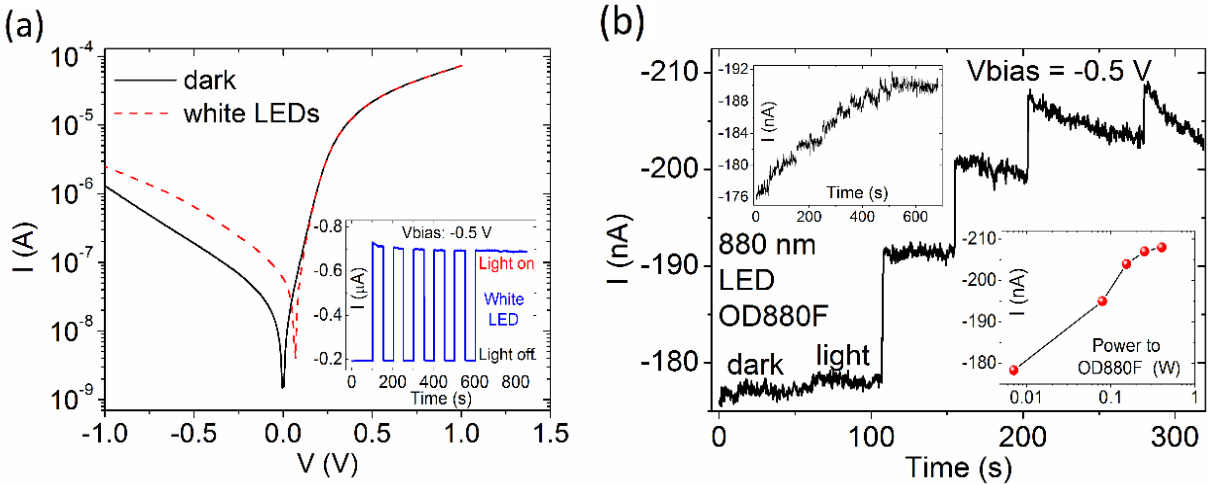

**Figure 7.** a) I-V characteristics of the graphene/Si-tip device in dark and under white LED illumination. The inset show the current at V=-0.5 V in a sequence of light on/off cycles. b)

Reverse current at V=-0.5 V under 880 nm IR irradiation, increased in time by stepwise varying the input electrical power to the emitting diode (OD880F). The top inset displays the monotonic increase of the photocurrent of the graphene/Si-tip device while the IR diode input power is changed in smaller steps. The bottom inset shows the current of the graphene/Si-tip vs. the IR diode input electrical power.

The high substrate doping reduces the space-charge region where most of the photoexcitation takes place and this should suppress the responsivity. This loss can be counterbalanced by the textured surface, which on the contrary favors multiple reflections and light absorption[64]. However, the high value of responsivity is rather a result of the peculiar device geometry. The tip-enhanced electric field, other than facilitating their separation, can provide photogenerated electron-hole pairs with enough kinetic energy to cause impact ionization and initiate charge multiplication, thus enabling device internal gain. The exponential increase of the photocurrent with reverse bias can be taken as signature of internal gain since smoother rise is usually observed when the photocurrent is due only to increased photon absorption in the bias-widened depletion layer. Augmented photocharge separation and multiplication is also expected in graphene, especially in the areas near the tips. We notice that very high responsivity of $10^7 \ A/W$ have been reported by Liu et al.[51] in more complex three-terminal Gr/Si devices with quantum gain due to photocarriers borrowed into graphene and reinvested several times in the external circuit during their lifetime. However, these devices require more complex circuitry than our multi-purpose two-terminal photodiode.

## 4. CONCLUSIONS

In conclusion, we have fabricated and extensively characterized a novel Gr/Si heterojunction obtained by CVD graphene transfer over a nanotip patterned Si-substrate. We have measured the relevant junction parameters and shown better performance of the Gr/Si-tip device with respect to its planar counterpart. Without adding barrier inhomogeneity, the tip geometry enables linear tuning of the Schottky barrier height, hence of the diode current, by a single applied bias, thus implementing a two terminal barristor. The textured surface improves light absorption and photocharge collection and enables internal gain through impact ionization leading to higher responsivity.

This study represent a step forward toward the integration of graphene into existing Si technology for new generation optoelectronic devices.

**AUTHOR INFORMATION**

**Corresponding Author**

*Antonio Di Bartolomeo e-mail: dibant@sa.infn.it .

**REFERENCES**

(1) An, X.; Liu, F.; Jung Y. J.; Kar S. Tunable Graphene–Silicon Heterojunctions for Ultrasensitive Photodetection. *Nano Lett.* **2013**, *13*, 909–916.

(2) Ferrari, A.; Bonaccorso, F.; Fal'ko, V.; Novoselov, K. S.; Roche, S.; Bøggild, P.; Borini, S.; Koppens, F. H. L.; Palermo, V.; Pugno, N. et al. Science and technology roadmap for graphene, related two dimensional crystals, and hybrid systems. *Nanoscale* **2015**, *7*, 4598-4810.

(3) Riazimehr, S.; Bablich, A.; Schneider, D.; Kataria, S.; Passi, V.; Yim, C.; Duesberg, G. S.; Lemme, M. C. Spectral Sensitivity of Graphene/Silicon Heterojunction Photodetectors. *Solid-State Electronics* **2016**, *115*, 207-212.

(4) An, X.; Liu, F.; Kar, S. Optimizing performance parameters of graphene–silicon and thin transparent graphite–silicon heterojunction solar cells. *Carbon* **2013**, *57*, 329–337.

(5) Behura, S. K.; Nayak, S.; Mukhopadhyay, I.; Jani O. Junction characteristics of chemically-derived graphene/p-Si heterojunction solar cell. *Carbon* **2014**, *67*, 766–774.

(6) Ruan, K.; Ding, K.; Wang, Y.; Diao, S.; Shao, Z.; Zhang, X.; Jie J. Flexible graphene/silicon heterojunction solar cells. *J. Mater. Chem. A* **2015**, *3*, 14370-14377.

(7) Kim, H.-Y. ; Lee, K.; McEvoy, N. ; Yim, C.;  Duesberg G. S. Chemically modulated graphene diodes. *Nano Lett.* **2013**, *13*, 2182–2188.

(8) Singh, A.; Uddin, A.; Sudarshan, T.; Koley G.  Tunable reverse-biased graphene/silicon heterojunction Schottky diode sensor. *Small* **2014**, 10, 1555-1565.

(9) Fattah, A.; Khatami, S.; Mayorga-Martinez, C. C.; Medina-Sánchez, M.; Baptista-Pires, L.; Merkoçi A. Graphene/silicon heterojunction schottky diode for vapors sensing using impedance spectroscopy. *Small* **2014**, 10, 4193–4199.

(10) Di Bartolomeo, A.; Giubileo, F.; Santandrea, S.; Romeo, F.; Citro, R; Schroeder T., Lupina G. Charge transfer and partial pinning at the contacts as the origin of a double dip in the transfer characteristics of graphene-based field-effect transistors. *Nanotechnology* **2011**, *22*, 275702 (8pp).

(11) Yang, H.; Heo, J.; Park, S.; Song, H.J.; Seo, D.H.; Byun, K.-E.; Kim, P.; Yoo, I.K. ; Chung, H.-J.; Kim, K. Graphene Barristor, a Triode Device with a Gate-Controlled Schottky Barrier. *Science* **2012**, *336*, 1140-1143.

(12) Mehr, W.; Dabrowski, J.; Scheytt, J. C.; Lippert, G.; Xie, Y.-H.; Lemme, M. C.; Ostling, M.; Lupina, G. Vertical graphene base transistor. *IEEE Electron Device Lett.*, **2012**, *33*, 691–693.

(13) Di Bartolomeo, A.; Santandrea, S.; Giubileo, F.; Romeo, F.; Petrosino, M.; Citro, R.; Barbara, P.; Lupina, G.; Schroeder, T.; Rubino A. Effect of back-gate on contact resistance and on channel conductance in graphene-based field-effect transistors. *Diamond Rel. Mat.* **2013**, *38*, 19–23.

(14) Vaziri, S.; Smith, A.D.; Östling, M.; Lupina, G.; Dabrowski, J.; Lippert, G.; Mehr, W.; Driussi, F.; Venica, S.; DiLecce, V.; Gnudi, A.; König, M.; Ruhl, G.; Belete, M.; Lemme, M. C. Going ballistic: Graphene hot electron transistors. *Solid State Commun.* **2015**, *224*, 64–75.

(15) Lupina, G.; Strobel, C.; Dabrowski, J.; Lippert, G.; Kitzmann, J.; Krause, H. M.; Wenger, Ch.; Lukosius, M.; Wolff, A.; Albert M.; Bartha, J. W. Plasma-enhanced chemical vapor deposition of amorphous Si on graphene. *Appl. Phys. Lett.* **2016**, *108*, 193105.

(16) Di Bartolomeo, A. Graphene Schottky diodes: An experimental review of the rectifying graphene/semiconductor heterojunction. *Physics Reports* **2016**, *606*, 1-58.

(17) Tongay, S.; Lemaitre, M.; Miao, X.; Gila, B.; Appleton, B.R.; Hebard, A.F. Rectification at Graphene-Semiconductor Interfaces: Zero-Gap Semiconductor-Based Diodes. *Phys. Rev. X* **2012**, *2*, 011002 (pp 9).

(18) Mehr, W.; Wolff, A.; Frankenfeld, H.; Skaloud, T.; Höppner, W.; Bugiel, E.; Lärz, J.; Hunger, B. Ultra Sharp Crystalline Silicon Tip Array Used as Field Emitter. *Microelectron. Eng.* **1996**, *30*, 395–398.

(19) Niu, G.; Capellini, G.; Lupina, G.; Niermann, T.; Salvalaglio M.; Marzegalli, A.; Schubert, M.A.; Zaumseil, P.; Krause, H.-M.; Skibitzki, O.; Lehmann, M.; Montalenti, F.; Xie, Y.-H.; Schroeder, T. Photodetection in Hybrid Single-Layer Graphene/Fully Coherent Germanium Island

Nanostructures Selectively Grown on Silicon Nanotip Patterns. *ACS Appl. Mat. and Interf.* **2016**, *8*, 2017-2026.

(20) Niu, G; Capellini, G; Schubert, M.A.; Niermann, T; Zaumseil, P; Katzer, J; Krause, H.-M.; Skibitzki, O; Lehmann, M; Xie Y.-H.; von Känel, H.; Schroeder T. Dislocation-free Ge Nano-crystals via Pattern Independent Selective Ge Heteroepitaxy on Si Nano-Tip Wafers. *Sci. Rep.* **2016**, *6*, 22709.

(21) Lupina, G.; Kitzmann, J.; Costina, I.; Lukosius, M.; Wenger, C.; Wolff, A.; Vaziri, S.; Ostling, M.; Pasternak, I.; Krajewska, A. et al. Residual metallic contamination of transferred chemical vapor deposited graphene. *ACS Nano* **2015**, *9*, 4776-4785.

(22) Lupina, G.; Lukosius, M.; Kitzmann, J.; Dabrowski, J.; Wolff, A.; Mehr, W. Nucleation and Growth of $HfO_2$ Layers on Graphene by Chemical Vapor Deposition. *Appl. Phys. Lett.* **2013**, *103*, 183116.

(23) Ferrari, A.C.; Basko D.M. Raman spectroscopy as a versatile tool for studying the properties of graphene. *Nat. Nanotech.* **2013**, *8*, 235-246.

(24) Lee, J.-K.; Park C.-S.; Kim, H. Sheet resistance variation of graphene grown on annealed and mechanically polished Cu films. *RSC Advances* **2014**, *4*, 62453-62456.

(25) Di Bartolomeo, A.; Giubileo, F.; Romeo, F.; Sabatino, P.; Carapella, G.; Iemmo, L.; Schroeder, T.; Lupina, G. Graphene field effect transistors with niobium contacts and asymmetric transfer characteristics. *Nanotechnology* **2015**, *26*, 475202.

(26) Tung, R.T. Recent advances in Schottky barrier concepts. *Mat. Sci. and Eng.* **2001**, R35, 1-138.

(27) Tung, R.T. The physics and chemistry of the Schottky barrier height. *Appl. Phys. Rev.* **2014**, *1*, 011304 (pp 1-54)

(28) Schroder, D.K. Semiconductor material and device characterization. *Wiley Interscience*, Hoboken, New Jersey, **2007**

(29) Sze, S.M.; Ng, K.K. Physics of semiconductor devices. *Wiley Interscience*, Hoboken, New Jersey, **2007**

(30) Cheung, S.K.; Cheung, N.W. Extraction of Schottky diode parameters from forward current-voltage characteristics. *Appl. Phys. Lett.* **1986**, *49*, 85.

(31) Olikh, O.Y. Review and test of methods for determination of the Schottky diode parameters. *J. Appl. Phys.* **2015**, 118, 024502.

(32) Werner, J.H. Schottky barrier and pn-junction*I/V* plots — Small signal evaluation J. *Appl. Phys.* **1988**, *47*, 291-300.

(33) Li S.S.; Thurber, W.R. The dopant density and temperature dependence of electron mobility and resistivity in *n*-type silicon. *Solid-State Electronics* **1977**, *20*, 609-616.

(34) Jacoboni, C.; Canali, C.; Ottaviani, G.; Alberigi Quaranta, A.; A review of some charge transport properties of silicon. *Solid-State Electronics* **1977**, *20*, 77-89.

(35) Swirhun, S.E.; Swanson, R.M. Temperature Dependence of Specific Contact Resistivity. *IEEE Electr. Dev. Lett.* **1986**, *7*, 155-157.

(36) Bolotin, K.I.; Sikes, K.J.; Hone, J.; Stormer, H.L.; Kim P. Temperature-Dependent Transport in Suspended Graphene. *Phys. Rev. Lett.* **2008**, *101*, 096802 (pp 1-4).

(37) Shao, Q.; Liu, G.; Teweldebrhan, D.; Balandin A.A. High-temperature quenching of electrical resistance in graphene interconnects. *Appl. Phys. Lett.* **2008**, *92*, 202108.

(38) Skákalová, V.; Kaiser, A.B.; Yoo, J.S.; Obergfell, D.; Roth, S. Correlation between resistance fluctuations and temperature dependence of conductivity in graphene. *Phys. Rev. B* **2009**, *80*, 153404 (pp 1-7).

(39) Heo, J.; Chung, H.J.; Lee, S.-H.; Yang, H.; Seo, D.H.; Shin, J.K.; Chung, U-I.; Seo, S.; Hwang, E.H.; Das Sarma S. Nonmonotonic temperature dependent transport in graphene grown by chemical vapor deposition. *Phys. Rev. B* **2011**, *84*, 035421 (7pp).

(40) Liu, Y.; Liu, Z.; Lew, W.S.; Wang Q.J. Temperature dependence of the electrical transport properties in few-layer graphene interconnects. *Nano. Res. Lett.* **2013**, *8*, 335 (7pp).

(41) Liu, Y.; Li, W.; Qi, M.; Li, X.; Zhou, Y.; Ren, Z. Study on temperature-dependent carrier transport for bilayer graphene. *Physica E* **2015**, *69*, 115–120.

(42) Fang, X.-Y.; Yu, X.-X.; Zheng, H.-M.; Jin, H.-B.; Wang, L.; Cao M.-S. Temperature-and thickness-dependent electrical conductivity of few-layer graphene and graphene nanosheets. *Phys. Lett. A* **2015**, *379*, 2245–2251.

(43) Werner J.H., Güttler H.H. Barrier inhomogeneities at Schottky contacts. *J. Appl. Phys.* **1991**, *69* 1522-1533.

(44) Parui, S.; Ruiter, R.; Zomer, P.J.; Wojtaszek, M.; Van Wees, B.J.; Banerjee, T. Temperature dependent transport characteristics of graphene/n-Si diodes, *J. Appl. Phys.* **2014**, *116*, 244505 (pp 5).

(45) Sinha, D.; Lee, J.U. Ideal Graphene/Silicon Schottky junction diodes, *Nano Lett.* **2014**, *14*, 4660-4664.

(46) Tomer, D.; Rajput, S.; Hudy, L.J.; Li C.H.; Li L. Inhomogeneity in barrier height at graphene/Si (GaAs) Schottky junctions. *Nanotechnology* **2015**, *26*, 215702 (7pp).

(47) Liang, S.-J.; Ang L.K. Revised diode equation for Ideal Graphene-Semiconductor Schottky Junction. *arXiv:1503.02758v2*, **2015**, pp 7.

(48) Liang, S.-J.; Ang L.K. Electron Thermionic Emission from Graphene and a Thermionic Energy Converter. *Phys. Rev. Applied* 2015, 3, 014002.

(49) Khurelbaatar, Z.; Kil, Y.-H.; Shim, K.-H.; Cho, H.; Kim, M.-J.; Kim, Y.-T.; Choi C.-J. Temperature Dependent Current Transport Mechanism in Graphene/Germanium Schottky Barrier Diode. *J. Semicond. Techn. Sci.* **2015**, *15*, 7-15.

(50) Chen, C.-C.; Aykol, M.; Chang, C.-C.; Levi, A.F.J.; Cronin, S.B. Graphene-Silicon Schottky Diodes, *Nano Lett.* **2011**, *11*, 1863-1867.

(51) Liu, F.; Kar, S. Quantum carrier reinvestment-induced ultrahigh and broadband photocurrent response in Graphene-Silicon Junctions, *ACS Nano* **2014**, *8*, 10270-10279.

(52) Zhong, H.; Xu, K.; Liu, Z.; Xu, G.; Shi, L.; Fan, Y.; Wang, J.; Ren, G.; Yang H. Charge transport mechanisms of graphene/semiconductor Schottky barriers: A theoretical and experimental study *J. Appl. Phys.* **2014**, *115*, 013701.

(53) Khomyakov, P.A.; Giovannetti, G.; Rusu, P.C.; Brocks, G.; van den Brink, J.; Kelly, P.J. First-principles study of the interaction and charge transfer between graphene and metals, *Phys. Rev. B* **2009**, *79*, 195425 (pp 12).

(54) Song, S.M.; Park, J.K.; Sul, O.J.; Cho, B.J. Determination of work function of graphene under a metal electrode and its role in contact resistance. *Nano Lett.* **2012**, *12*, 3887-3892.

(55) Xu, Y.; He, K.T.; Schmucker, S.W.; Guo, Z.; Koepke, J.C.; Wood, J.D.; Lyding, J.W.; Aluru N.R. Inducing Electronic Changes in Graphene through Silicon (100) Substrate Modification. *Nano Lett.*, **2011**, *11*, 2735–2742.

(56) Lin, Y-J; Zeng J.-J. Determination of Schottky barrier heights and Fermi-level unpinning at the graphene/n-type Si interfaces by X-ray photoelectron spectroscopy and Kelvin probe. *Appl. Surf. Science* **2014**, *322*, 225-229.

(57) Rideout, V.L. A review of the theory, technology and applications of Metal-Semiconductor rectifiers, *Thin Solid Films* **1978**, *48*, 261-291.

(58) Chang, W.; Shih, C.-H.; Luo, Y.-X.; Wu, W.-F.; Lien C. Drain-induced Schottky barrier source-side hot carriers and its application to program local bits of nanowire charge-trapping memories. *Jap. J. Appl. Phys.* **2014**, *53*, 094001.

(59) Mao, L.-F.; Ning, H.; Huo, Z.-L.; Wang, J.-Y. Physical Modeling of Gate-Controlled Schottky Barrier Lowering of Metal-Graphene Contacts in Top-Gated Graphene Field-Effect Transistors *Sci. Rep.* **2015**, *5*, 18307 (pp. 1-11)

(60) Echtermeyer, T. J.; Britnell, L.; Jasnos, P. K.; Lombardo, A.; Gorbachev, R.V.; Grigorenko, A.N.; Geim, A.K.; Ferrari, A.C.; Novoselov, K.S. Strong Plasmonic enhancement of photovoltage in graphene. *Nat. Commun.* **2011**, *2*, 458 (pp 1-5).

(61) Furchi, M.; Urich, A.; Pospischil, A.; Lilley, G.; Unterrainer, K.; Detz, H.; Klang, P.; Andrews, A. M.; Schrenk, W.; Strasser, G.; Mueller, T. Microcavity-Integrated Graphene Photodetector. *Nano Lett.* **2012**, *12*, 2773−2777.

(62) Amirmazlaghani, M.; Raissi, F.; Habibpour, O.; Vukusic, J.; Stake, J. Graphene-Si Schottky IR Detector. *IEEE J. Quant. Electr.* **2013**, *49*, 589-594.

(63) Lv, P.; Zhang, Xiujuan; Zhang, Xiwei; Deng, W. ; Jie J. High-Sensitivity and Fast-Response Graphene/Crystalline Silicon Schottky Junction-Based Near-IR Photodetectors. *IEEE Elect. Dev. Lett.* **2013**, 34, 1337-1339.

(64) Khan, F; Baek, S.-H.; Kaur, J.; Fareed, I.; Mobin A.; Kim J.H. Paraboloid Structured Silicon Surface for Enhanced Light Absorption: Experimental and Simulative Investigations. *Nano. Res. Lett.* **2015**, 10, 376-384.

**Table of contents graphic**

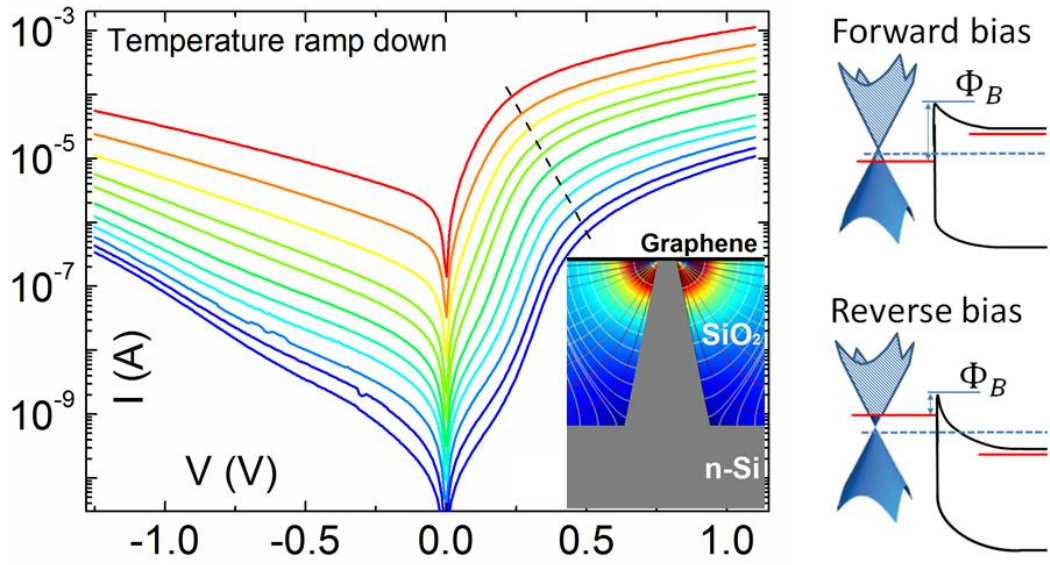